\documentclass[final,5p,times,twocolumn]{elsarticle}

\usepackage{hyperref}
\usepackage{graphicx}
\usepackage{amssymb}

\begin{document}

\begin{frontmatter}

\title{On particle collisions near Kerr's black holes}

\author[label1,label2]{Andrey A. Grib} %% \corref{cor1}}
\ead{andrei\_grib@mail.ru}

\author[label1,label3]{Yuri V. Pavlov}
\ead{yuri.pavlov@mail.ru}

\address[label1]{A. Friedmann Laboratory for Theoretical Physics,
30/32 Griboedov can., St.\,Petersburg, 191023, Russia}

\address[label2]{Theoretical Physics and Astronomy Department, The
Herzen  University, 48, Moika, St.\,Petersburg, 191186, Russia}

\address[label3]{Institute of Mechanical Engineering, RAS,
61 Bolshoy, V.O., St.\,Petersburg, 199178, Russia}

%%\cortext[cor1]{Corresponding author.}

\begin{abstract}
    Scattering of particles in the gravitational field
of rotating black holes is considered.
    Expressions for scattering energy of particles in the centre of mass
system are obtained.
    It is shown that scattering energy of particles in the centre of mass
system can obtain very large values not only for extremal black holes
but also for nonextremal ones.
    It is shown that for realizing of the collisions with infinite energy
one needs the infinite interval not only of the coordinate time but
also the infinite interval of the proper time of the free falling particle.

\end{abstract}

\begin{keyword}
black holes \sep particle collisions

\PACS 04.70.-s, 04.70.Bw, 97.60.Lf

\end{keyword}

\end{frontmatter}

%% main text
%%%%%%%%%%%%%%%%%%%%%%%%%%%%%%%%%%%%%%%%%%%%%%%%%%%%%%%%%%%%%%%%%%%%%%
\section{Introduction}
\label{1secBHColl}
%%%%%%%%%%%%%%%%%%%%%%%%%%%%%%%%%%%%%%%%%%%%%%%%%%%%%%%%%%%%%%%%%%%%%%

    In~\cite{GribPavlov2007AGN} we put the hypothesis that
active galactic nuclei can be the sources of ultrahigh energy particles
in cosmic rays observed recently by the AUGER group
(see~\cite{Auger07}) due to the processes of converting dark matter
formed by superheavy neutral particles into visible particles --- quarks,
leptons (neutrinos), photons.
    If active galactic nuclei are rotating black holes then
in~\cite{GribPavlov2007AGN} we discussed the idea that
``This black hole acts as a cosmic supercollider in which superheavy
particles of dark matter are accelerated close to the horizon to
the Grand Unification energies and can be scattering in collisions.''
    It was also shown~\cite{GribPavlov2008KLGN} that in
Penrose process~\cite{Penrose69}
dark matter particle can decay on two particles, one with the negative
energy, the other with the positive one and particles of very high
energy of the Grand Unification order can escape the black hole.
    Then these particles due to interaction with photons close to
the black hole will loose energy analogously up to
the Greisen-Zatsepin-Kuzmin limit in cosmology~\cite{GZK}.

    First calculations of the scattering of particles in the ergosphere
of the rotating black hole, taking into account the Penrose process, with
the result that particles with high energy can escape the black hole, were
made in~\cite{PiranShahamKatz75,PiranShaham77}.
    Recently in~\cite{BanadosSilkWest09} it was shown that for the rotating
black hole (if it is the critical one) the energy of scattering
is unlimited.
    The result of~\cite{BanadosSilkWest09} was criticized
in~\cite{BertiCardosoGPS09} in the sense that it does not
occur in nature.
    The authors of~\cite{BertiCardosoGPS09} claimed that
if the black hole is not a critical rotating black hole so that its
dimensionless angular momentum $A \ne 1$ but $ A=0.998$ then
the energy is limited.
    In this paper we show that the energy of scattering in the centre of mass
system can be still unlimited in the cases of multiple scattering.

    In the part~\ref{2secBHColl}  we calculate the energy of collisions
in the centre mass system for the particles falling onto a rotating black hole.
    For the case of the nonrelativistic at infinity particles we reproduce
the results of~\cite{BanadosSilkWest09}.
    In the part~\ref{3secBHnonextr} we consider the case of the nonextremal
black holes and show that in some cases (multiple scattering) the results
of~\cite{BertiCardosoGPS09} on the limitations of
the scattering energy for the nonextremal black holes are not valid.
    In the part~\ref{3secEnUFN} the collisions inside black holes
are considered.
    The limiting formulas are obtained and it is shown
that the collisions with infinite energy can not be realized.
    In the part~\ref{31subsecEnUFN} we show that
for realizing of the collisions with infinite energy near events horizon
of  black holes one needs the infinite interval of as coordinate as proper
time of the free falling particle.

    The system of units $G=c=1$ is used in the paper.

%%%%%%%%%%%%%%%% Section "Energy of Collision in Black Holes" %%%%%%%%
\section{The energy of collisions in the field of Kerr's black hole}
\label{2secBHColl}

    Let us consider particles falling on the rotating chargeless black hole.
    The Kerr's metric of the rotating black hole in Boyer--Lindquist
coordinates has the form                    %% R.H.Boyer, R.W.Lindquist
    \begin{eqnarray}
d s^2 = d t^2 -
\frac{2 M r \, ( d t - a \sin^2 \! \theta\, d \varphi )^2}{r^2 + a^2 \cos^2
\! \theta } \hspace{33pt}
\nonumber \\
- (r^2 + a^2 \cos^2 \! \theta ) \left( \frac{d r^2}{\Delta} + d \theta^2 \right)
- (r^2 + a^2) \sin^2 \! \theta\, d \varphi^2, \ \
\label{Kerr}
\end{eqnarray}
    where
    \begin{equation} \label{Delta}
\Delta = r^2 - 2 M r + a^2,
\end{equation}
    $M$ is the mass of the black hole, $J=aM$ is angular momentum.
    In the case $a=0$ the metric~(\ref{Kerr}) describes the static chargeless
black hole in Schwarzschild coordinates.
    The event horizon for the Kerr's black hole corresponds to the value
    \begin{equation}
r = r_H \equiv M + \sqrt{M^2 - a^2} \,.
\label{Hor}
\end{equation}
    The Cauchy horizon is
    \begin{equation}
r = r_C \equiv M - \sqrt{M^2 - a^2} \,. \label{HorCau}
\end{equation}
    The surface called ``the static limit'' is defined by the expression
     \begin{equation}
r = r_0 \equiv M + \sqrt{M^2 - a^2 \cos^2 \! \theta} \,.
\label{PrSt}
\end{equation}
    The region of space-time between the horizon and the static limit is
ergosphere.

    For equatorial ($\theta=\pi/2$) geodesics in Kerr's metric~(\ref{Kerr}) one
obtains (\cite{Chandrasekhar}, \S\,61):
    \begin{equation} \label{geodKerr1}
\frac{d t}{d \tau} = \frac{1}{\Delta} \left[ \left(
r^2 + a^2 + \frac{2 M a^2}{r} \right) \varepsilon - \frac{2 M a}{r} L \right],
\end{equation}
    \begin{equation}
\frac{d \varphi}{d \tau} = \frac{1}{\Delta} \left[ \frac{2 M a}{r}\,
\varepsilon + \left( 1 - \frac{2 M}{r} \right)\! L \right],
\label{geodKerr2}
\end{equation}
    \begin{equation} \label{geodKerr3}
\left( \frac{d r}{d \tau} \right)^2 = \varepsilon^2 +
\frac{2 M}{r^3} \, (a \varepsilon - L)^2 +
\frac{a^2 \varepsilon^2 - L^2}{r^2} - \frac{\Delta}{r^2}\, \delta_1 \,,
\end{equation}
    where
    $\delta_1 = 1 $ for timelike geodesics
($\delta_1 = 0 $ for isotropic geodesics),
$\tau$ is the proper time of the moving particle,
$\varepsilon={\rm const} $ is the specific energy:
the particle with rest mass~$m$ has the energy $\varepsilon m $ in the
gravitational field~(\ref{Kerr});
$ L m = {\rm const} $ is the angular momentum of the particle relative
to the axis orthogonal to the plane of movement.

%%%%%%%%%%%%%%%%%%%%%%%%%%%%%%%%%%%%%%%%%%%%%%%%%%
    Let us find the energy $E_{\rm c.m.}$ in the centre of mass system
of two colliding particles with the same rest~$m$ in arbitrary gravitational
field.
    It can be obtained from
    \begin{equation} \label{SCM}
\left( E_{\rm c.m.}, 0\,,0\,,0\, \right) =
m u^i_{(1)} + m u^i_{(2)}\,,
\end{equation}
    where $u^i=dx^i/ds$.
    Taking the squared~(\ref{SCM}) and due to $u^i u_i=1$ one obtains
    \begin{equation} \label{SCM2}
E_{\rm c.m.} = m \sqrt{2}\, \sqrt{1+ u_{(1)}^i u_{(2) i}} \,.
\end{equation}
    The scalar product does not depend on the choice of the coordinate frame
so~(\ref{SCM2}) is valid in an arbitrary coordinate system and for arbitrary
gravitational field.
%%%%%%%%%%%%%%%%%%%%%%%%%%%%%%%%%%%%%%%%%%%%%%%%%%

    We denote~$x=r/M$, \ $ A=a/M$, \ $ l_n=L_n/M$,
\ $ \Delta_x = x^2 - 2 x + A^2 $     and
    \begin{equation} \label{DenKBHxhc}
x_H = 1 + \sqrt{1 - A^2}, \ \ \
x_C = 1 - \sqrt{1 - A^2}.
\end{equation}
    For the energy in the centre
of mass frame of two colliding particles with
$\varepsilon_1 = \varepsilon_2 = \varepsilon $ and
angular momenta $L_1, \, L_2$,
which are moving in Kerr's metric one obtains
using~(\ref{Kerr}), (\ref{geodKerr1})--(\ref{geodKerr3}), (\ref{SCM2}):
    \begin{eqnarray}
\frac{E_{\rm c.m.}^2}{2\, m^2} = 1 - \varepsilon^2 +
\frac{1}{x \Delta_x} \Biggl[ 2 \varepsilon^2 x^2 (x-1) + l_1 l_2 (2-x)
\nonumber \\
+\, 2 \varepsilon^2 A^2 (x+1) - 2 \varepsilon A (l_1 +l_2 )
\nonumber \\
- \sqrt{ 2 \varepsilon^2 x^2 + 2 (l_1 - \varepsilon  A )^2 -l_1^2 x
+ (\varepsilon^2 - 1 ) x \Delta_x }
\nonumber  \\
\times \sqrt{ 2 \varepsilon^2 x^2 + 2 (l_2 - \varepsilon  A )^2 -l_2^2 x
+ (\varepsilon^2 - 1 ) x \Delta_x }\, \Biggr]\,.
\label{KerrL1L2}
\end{eqnarray}
    It corresponds to result in~\cite{BanadosSilkWest09} for
the case $ \varepsilon=1$.

    To find the limit $r \to r_H$ for the black hole with a given angular
momentum~$A$ one must take in~(\ref{KerrL1L2}) $x = x_H + \alpha$
with $\alpha \to 0 $ and do calculations up to the order~$\alpha^2$.
    Taking into account $ A^2 = x_H x_C$, $x_H + x_C=2$, after resolution
of uncertainties in the limit $\alpha \to 0 $ one obtains
    \begin{equation} \label{KerrLimA}
\frac{E_{\rm c.m.}(r \to r_H) }{2 m}
= \sqrt{1 +
\frac{(l_1-l_2)^2 (4+ l_H^2)}{16 (l_H - l_1) (l_H - l_2)}} \,,
\end{equation}
    where
    \begin{equation} \label{KerrlH}
l_H = \frac{2 \varepsilon x_H}{A} =  \frac{2 \varepsilon}{A}
\left( 1 + \sqrt{1-A^2} \, \right).
\end{equation}
is the limiting value of the angular momentum of the particle close
to the horizon of the black hole.
    It can be obtained from the condition
of positive derivative in~(\ref{geodKerr1}) $dt /d \tau > 0$,
i.e. going ``forward'' in time:
    \begin{equation} \label{KerrLehxa}
l < l_H \left( 1 + \frac{x_H +1}{2}\, \alpha \right) + o(\alpha), \ \
x =x_H+ \alpha .
\end{equation}
    So close to the horizon one has the condition
$l \le l_H $.

    For the extremal black hole $A=x_H=1$, $ l_H=2 \varepsilon $ and
the expression~(\ref{KerrLimA}) takes the form
    \begin{equation} \label{KerrLimAExtr}
\frac{E_{\rm c.m.}(r \to r_H) }{2 m}
= \sqrt{1 + \frac{ 1 + \varepsilon^2}{4}
\frac{(l_1-l_2)^2}{(2 \varepsilon - l_1) ( 2 \varepsilon - l_2)}} \,,
\end{equation}
    is divergent when the dimensionless angular momentum of
one of the falling into the black hole particles $ l= 2 \varepsilon$.
    The scattering energy in the centre of mass system is increasing without
limit
(for case $\varepsilon=1$ it was established in~\cite{BanadosSilkWest09}).
    For example, if $l_1=l_H $ then  one obtains from Eq.~(\ref{KerrL1L2})
    \begin{equation} \label{KerrLimAExxx}
\frac{E_{\rm c.m.}(x) }{2 m} \approx
\sqrt{ \frac{l_H- l_2}{x-1} \left( \varepsilon - \frac{\sqrt{3 \varepsilon^2 -1}}{2}
\right) }\,, \ \ \ \ x \to 1 .
\end{equation}

    Can one get the unlimited high energy of this scattering energy for
the case of nonextremal black hole?

%%%% Section "The energy of collisions for nonextremal black hole" %%%
\section{The energy of collisions for nonextremal black hole}
\label{3secBHnonextr}

    In this section we consider the case $ \varepsilon = 1$,
when the particles falling into the black hole are
nonrelativistic at infinity.
    Formula~(\ref{geodKerr3}) leads to limitations on the possible values
of the angular momentum of falling particles:
the massive particle free falling in the black hole with dimensionless
angular momentum~$A$ to achieve the horizon of the black hole must have
angular momentum from the interval
    \begin{equation} \label{geodKerr5}
- 2 \left( 1 + \sqrt{1+A}\, \right) =l_L \le l
\le l_R = 2 \left( 1 + \sqrt{1-A}\, \right).
\end{equation}

    Putting the limiting values of angular momenta $l_L, l_R$
into the formula~(\ref{KerrLimA}) one obtains the maximal values
of the collision energy of particles freely falling from infinity
    \begin{eqnarray} \label{KerrLimAMax}
E_{\rm c.m.}^{\, \rm inf}(r \to r_H) =
\frac{2m}{\sqrt[4]{1-A^2}} \hspace{44pt}
\\
\times \sqrt{\frac{1-A^2+\left( 1+ \sqrt{1+A} +\sqrt{1-A} \right)^2}
{1+\sqrt{1-A^2}} }\,.
\nonumber
\end{eqnarray}
    For $A=1-\epsilon$ with $\epsilon \to 0$ formula~(\ref{KerrLimAMax})
gives:
    \begin{equation} \label{KerrimAE}
E_{\rm c.m.}^{\, \rm inf}(r \to r_H)
\sim 2 \left( 2^{1/4}+2^{-1/4} \right) \frac{m}{\epsilon^{1/4}} \,.
%%\approx \frac{m \cdot 4,06}{\epsilon^{1/4} } \,. %% 4.0602070605...
\end{equation}
    So even for values close to the extremal $A=1$ of the rotating black hole
$E_{\rm c.m.}^{\, \rm inf}/ m$ can be not very large as mentioned
in~\cite{BertiCardosoGPS09}.
    So for $A_{\rm max} =0.998 $ considered as the maximal possible
dimensionless angular momentum of the astrophysical black holes
(see~\cite{Thorne74}), from~(\ref{KerrLimAMax}) one obtains
$ E_{\rm c.m.}^{\, \rm max} /m \approx 18.97 $.

    Does it mean that in real processes of particle scattering in
the vicinity of the rotating nonextremal black holes the scattering energy
is limited so that no Grand Unification or even Planckean energies can
be obtained?
    Let us show that the answer is no!
    If one takes into account the possibility of multiple scattering so that
the particle falling from infinity on the black hole with some fixed
angular momentum changes its momentum in the result of interaction with
particles in the accreting disc and after this is again scattering close to
the horizon then the scattering energy can be unlimited.

\begin{figure}[ht]
    \includegraphics[width=86mm]{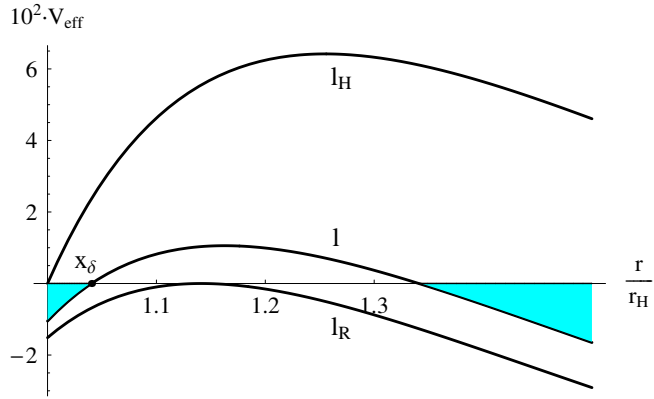}
\caption{ \label{Veff} The effective potential for $A=0.95$ and
$l_R \approx 2.45 $, \ %%2.44721}
$ l=2.5 $, \
$l_H \approx 2.76 $. %%2.76263}.
Allowed zones for \, $ l=2.5 $ are shown by the green color.
}
\end{figure}

    From~(\ref{geodKerr3}) one can obtain the permitted interval in~$r$ for
particles with $ \varepsilon = 1 $ and angular momentum $l = l_H - \delta $.
    To do this one must put the left hand side of~(\ref{geodKerr3})
to zero and find the root.
    In the second order in~$\delta$ close to the horizon one obtains
    \begin{equation} \label{KerrIntl}
l = l_H - \delta \ \ \Rightarrow \ \ \
x < x_\delta \approx x_H + \frac{\delta^2 x_C^2}{4 x_H \sqrt{1-A^2} } \,.
\end{equation}
    The effective potential for the case $ \varepsilon = 1 $
defined by the right hand side of~(\ref{geodKerr3})
    \begin{equation} \label{KerrVeff}
V_{\rm eff}(x,l) = - \frac{1}{2} \left( \frac{dr}{d \tau} \right)^2=
- \frac{1}{x} + \frac{l^2}{2x^2} - \frac{(A-l)^2}{x^3}
\end{equation}
(see, for example, Fig.~\ref{Veff})
%%%%%%%%%%%%%%%%%%%%%%%%%%%%%%%%%%%%%%%%%%%%%%%%%%
leads to the following behaviour of the particle.
    If the particle goes from infinity to the black hole it can achieve
the horizon if the inequality~(\ref{geodKerr5}) is valid.
    However the scattering energy in the centre of mass frame given
by~(\ref{KerrLimAMax}) is not large.
    But if the particle is going not from the infinity but
from some distance defined by~(\ref{KerrIntl}) then due to the form of
the potential it can have values of $l=l_H - \delta$
large than $l_R$ and fall on the horizon.
    If the particle falling from infinity with $ l \le l_R$ arrives
to the region defined by~(\ref{KerrIntl}) and here it interacts with other
particles of the accretion disc or it decays into a lighter particle which
gets an increased angular momentum $l_1 = l_H - \delta $,
then due to~(\ref{KerrLimA}) the scattering energy in the centre of mass
system is
    \begin{equation} \label{KerrInEn}
E_{\rm c.m.} \approx \frac{m}{\sqrt{\delta}} \,
\sqrt{ \frac{ 2(l_H - l_2) }{ 1- \sqrt{1 - A^2} }}
\end{equation}
    and it increases without limit for $\delta \to 0$.
    For $A_{\rm max} =0.998 $ and $l_2=l_L$, \
$ E_{\rm c.m.} \approx 3.85 m / \sqrt{\delta} $.  %% 3.85414

    Note that for rapidly rotating black holes $A= 1 - \epsilon$
the difference between $l_H$ and $l_R$ is not large
    \begin{eqnarray}
l_H - l_R &=& 2 \frac{\sqrt{1-A}}{A} \left( \sqrt{1-A} + \sqrt{1+A} -A \right)
\nonumber \\
&\approx& 2 (\sqrt{2}-1) \sqrt{\epsilon}\,, \ \ \ \epsilon \to 0 \,.
\label{KerrInDLR}
\end{eqnarray}
    For $A_{\rm max} =0.998 $, \ $ l_H - l_R \approx 0.04$   %% 0.0412465, 0.0370484
so the possibility of getting small additional angular momentum in
interaction close to the horizon seems much probable.
    The probability of multiple scattering in the accretion disc depends on
its particle density and is large for large density.
    Second scattering surely can be not on one trajectory
(which is improbable) with fixed angular
momentum but on all trajectories with angular momenta from the
interval~(\ref{geodKerr5}).

    Here we consider the model when the gravitation of the accretion disc
is treated as some perturbation much smaller than the gravitation of
the black hole so it is not taken into account.
    One must also mention that ``particles'' are considered as elementary
particles and not macroscopic bodies.
    So their ``large'' energy is limited by a Planckean value and we neglect
back reaction of it on the Kerr's metric of the macroscopic black hole.
    Electromagnetic and gravitational radiation of the particle surely
can change the picture but one needs exact calculations to see
what will be the balance.

%%%%%%%%%%%%%%%%%%%%%%%%%%%%%%%%%%%%%%%%%%%%%%%%%%%%%%%%%%%%%%%%%%%%%%
{\section{Collision of particles inside the rotating black hole}
\label{3secEnUFN}
}

    As one can see from formula~(\ref{KerrL1L2}) the infinite value
of the collision energy in the centre of mass system can be obtained inside
the horizon of the black hole on the Cauchy horizon~(\ref{HorCau}).
    Indeed, the zeroes of the denominator
in~(\ref{KerrL1L2}): $ x=x_H, \ x=x_C, \ x=0$.

    Let us find the expression for the collision energy for $x \to x_C$.
    Denote
    \begin{equation} \label{KerrlC}
l_C = \frac{2 \varepsilon x_C}{A} =
\frac{2 \varepsilon}{A} \left( 1 - \sqrt{1-A^2} \, \right).
\end{equation}
    Note that for $\varepsilon=1$
the Cauchy horizon can be crossed by the free falling from
the infinity particle under the same conditions on the angular
momentum~(\ref{geodKerr5}) as in case of the event horizon
and \ $ l_L < l_C \le l_R \le l_H$.

    To find the limit $r \to r_C$ for the black hole with a given angular
momentum~$A$ one must take in~(\ref{KerrL1L2}) $x = x_C + \alpha$
and do calculations with $\alpha \to 0 $ .
    The limiting energy has three different expressions depending on the values
of angular momenta.
    If
    \begin{equation} \label{lvnu1}
(l_1 - l_C) (l_2 - l_C) > 0 \,,
\end{equation}
i.e. $ l_1, l_2 $ are either both larger than $l_C$, or both smaller than $l_C$,
then
    \begin{equation} \label{EcmVnu1}
\frac{E_{\rm c.m.}(r \to r_C) }{2 m}
= \sqrt{1 +
\frac{(l_1-l_2)^2 (4+ l_C^2)}{16 (l_C - l_1) (l_C - l_2)}} \,.
\end{equation}
    This formula is similar to~(\ref{KerrLimA})
if everywhere $H \leftrightarrow C$. \
    If
    \begin{equation} \label{lvnu00}
(l_1 - l_C) (l_2 - l_C) = 0 \,,
\end{equation}
for example, $l_1=l_C$, then
    \begin{equation} \label{EcmVnu00}
\frac{E_{\rm c.m.}}{2 m} \approx
\sqrt[4]{\frac{(l_2-l_C)^2 (\varepsilon^2 x_C + x_H)}{
4 x_C (x_H-x_C) (x-x_C)}}\,, \  \ x \to x_C.
\end{equation}
    If
    \begin{equation} \label{lvnu2}
(l_1 - l_C) (l_2 - l_C) < 0 \,,
\end{equation}
    i.e. $ l_2 \in (l_L, \, l_C ), \ \ l_1 \in (l_C, \, l_R )$
(or the opposite), then
    \begin{equation} \label{EcmVnu2}
\frac{E_{\rm c.m.}}{2 m} \approx
\sqrt{\frac{ x_H (l_1 - l_C)(l_C - l_2)}{
x_C (x_H - x_C) (x - x_C)}}\,, \  \ x \to x_C.
\end{equation}

    It is seem that the limits of~(\ref{EcmVnu00}) and~(\ref{EcmVnu2})
is infinite for all values of angular
momenta $l_1,l_2$~(\ref{lvnu00}) and~(\ref{lvnu2}).
    However, from Eq.~(\ref{geodKerr1}) we can see
    \begin{equation} \label{EcmVnuLimll}
\frac{d t}{d \tau}(x \to x_C + 0 ) = \left\{
\begin{array}{ll}
+ \infty, & \ {\rm if} \ \ l > l_C\,, \\
- \infty, & \ {\rm if} \ \ l < l_C\,.
\end{array} \right.
\end{equation}
    That is why the collisions with infinite energy can not be
realized (see also~\cite{Lake10}).

%%%%%%%%%%%%%%%%%%%%%%%%%%%%%%%%%%%%%%%%%%%%%%%%%%%%%%%%%%%%%%%%%%%%%%
{\section{The time of movement before the collision with unbounded energy}
\label{31subsecEnUFN}
}

    Let us show that in order to get the unboundedly growing energy one must
have the time interval from the beginning of the falling inside the black
hole to the moment of collision also growing infinitely.
    This is connected with the fact of infinity of coordinate interval of
time needed for a freely particle to cross the horizon of the black hole
(for Schwarzschild metric this question was considered
in~\cite{GribPavlov2008UFN}).

    From Eq.~(\ref{KerrL1L2}) one can see that the collision energy can get
large values in the centre of mass system only if the collision occurs close
to the horizon.
    Unboundedly large energy of collisions outside of a black hole
is possible only for collisions on horizon $x \to x_H$
(see~(\ref{KerrLimAExxx})).

    From equation of the equatorial geodesic~(\ref{geodKerr1}),
(\ref{geodKerr3}) for a particle with dimensionless angular momentum~$l$
and specific energy $\varepsilon=1$ (i.e. the particle is non relativistic
at infinity) falling on the black hole with dimensionless angular momentum~$A$
one obtains
    \begin{equation} \label{Kerrdrdt}
\frac{d r}{d t} =  - \frac{(x- x_H) (x- x_C)}{\sqrt{x}} \,
\frac{\sqrt{ 2 x^2 - l^2 x + 2 (A -l)^2} }{x^3 +A^2 x + 2 A (A-l) }\,.
\end{equation}
    So the coordinate time (proper time of the observer at rest far from
the black hole) of the particle falling from some point $ r_0 = x_0 M $
to the point $r_f = x_f M > r_H$ is equal to
    \begin{equation} \label{KerrDelt}
\Delta t = M \int \limits_{x_f}^{x_0}
\frac{\sqrt{x} \left(x^3 +A^2 x + 2 A (A-l) \right)\, d x}{(x- x_H) (x- x_C)
\sqrt{ 2 x^2 - l^2 x + 2 (A -l)^2}} \,.
\end{equation}
    In case of the extremal rotating black hole ($A=1$, $x_C =x_H =1 $)
and the limiting value of the angular momentum $l=2$
the integral~(\ref{KerrDelt}) is equal to
    \begin{equation} \label{KerrDeltA1l2}
\Delta t = \frac{M}{\sqrt{2}} \left. \left(
\frac{2 \sqrt{x}\, (x^2 + 8 x - 15) }{ 3(x-1)} + 5 \ln
\frac{\sqrt{x} - 1}{\sqrt{x} + 1} \, \right)\, \right|_{x_f}^{x_0}
\end{equation}
    and it diverges as $ (x_f - 1)^{-1} $ for $ x_f \to 1 $.

    For the interval of proper time of the free falling to the black hole
particle one obtains from~(\ref{geodKerr3})
    \begin{equation} \label{KerrDeltau}
\Delta \tau = M \int \limits_{x_f}^{x_0}
\frac{\sqrt{x^3}\, d x}{
\sqrt{ 2 x^2 - l^2 x + 2 (A -l)^2}} \,.
\end{equation}
    For $A=1$, $l=2$ the integral~(\ref{KerrDeltau})) is equal to
    \begin{equation} \label{KerrDeltauA1l2}
\Delta \tau = \frac{M}{3 \sqrt{2}} \left. \left(
2 \sqrt{x}\, (3+ x) + 3 \ln \frac{\sqrt{x} - 1}{\sqrt{x} + 1}
\, \right)\, \right|_{x_f}^{x_0}
\end{equation}
    and it diverges logarithmically when $ x_f \to 1 $.
    So to get the collision with infinite energy one needs the infinite
interval not only of the coordinate time but also the infinite interval
of the proper time of the free falling particle.

%%%%%%%%%%%%%%%%%%%%%%%%%%%%%%%%%%%%%%%%%%%%%%%%%%%%%%%%%%%%%%%%%%%%%%

\end{document}